\documentstyle[12pt,epsf]{article}
\begin{document}
\newcommand{\frhalf}{\frac{1}{2}}
\newcommand{\sign}{\mbox{sign}}
\newcommand{\cl}{{$\cal L$}}
\newcommand{\mcl}{{\cal L}}
\newcommand{\ek}{\mbox{$\epsilon_{k_1}$}}
\newcommand{\ekp}{\mbox{$\epsilon_{k_2}$}}
\newcommand{\bk}{\mbox{$b_{k_1}$}}
\newcommand{\bkdag}{\mbox{$ b_{k_1}^\dagger$}}
\newcommand{\bkp}{\mbox{$b_{k_2}$}}
\newcommand{\bkpdag}{\mbox{$ b_{k_2}^\dagger$}}
\newcommand{\beq}{\begin{equation}}
\newcommand{\eeq}{\end{equation}}
\newcommand{\delth}{ \frac{d}{d \th}}
\newcommand{\th}{\theta}
\newcommand{\bbl}{ [ }
\newcommand{ \bbr} { ]_k }
\title{  \bf Superconductivity from Repulsion: A Variational View}
\author{{\bf B Sriram Shastry}\\
 Indian Institute of Science\\
 Bangalore 560012, INDIA}
\maketitle
\date{March 22, 1999}


\begin{abstract}
This work present a new class of variational wave functions for fermi systems in
any dimension. These wave functions introduce correlations between Cooper pairs
in different momentum states and  the relevant correlations can be computed
analytically. 
At half filling we have a ground state with critical  superconducting correlations,
 that causes  negligible increase of the kinetic energy. 
We find large enhancements in a Cooper pair correlation function
caused purely by the interplay between the uncertainty principle, 
repulsion and the proximity of half filling. This
is surprising since there is no accompanying signature in usual charge and
spin response functions, and typifies a novel kind of many body
cooperative behaviour.
\end{abstract}

\newpage

{\em \bf Introduction:}
Variational wave functions have played an important part in our understanding
of several phenomena in  quantum manybody systems. The role of the BCS \cite{bcs} wave function in superconductivity,  Jastrow-type functions for 
superfluidity \cite{jastrow}, the Laughlin  function
in the Fractional Quantum Hall problem\cite{laughlin}, and the Gutzwiller
wavefunction \cite{gutzwiller}
in strongly correlated fermi systems have been recognized to be of
fundamental importance. 

 This work  presents a new variational wave function 
and some variants of it, that promise to be
of  interest in the topical problem of superconductivity arising from repulsion.
We have recently argued on the basis of certain inequalities\cite{shas}, that projecting out s-wave Cooper pairs in a fermi system {\em on a lattice } would lead to enhanced extended s-wave type pairing fluctuations near half filling.
These are
expected to lead to
superconductivity at precisely half filling
for a class of Hubbard-type models.
In brief the argument involves the recognition that s-wave and extended s-wave pairing are 
canonically conjugate in the sense of the uncertainty principle, which
ultimately drives the instability towards superconductivity
near half filling. This conjugacy arises since
 the s wave and the extended s wave Cooper order parameters
$B, A$ ( defined below)
satisfy the commutation relation $[B, A^\dagger]= 2 T$,  where the kinetic energy
operator $T$ plays a benign  role, similar to that of the 
number operator $\hat{N}$
as in other familiar contexts, such as the conductivity sum rule on 
the lattice. From the uncertainty principle, it follows\cite{shas}
that the fluctuation $<A^\dagger A + A A^\dagger>$ is bounded from below by
$4 <T>^2/< B^\dagger B + B B^\dagger>$.
The special role played by half filling is due to the fact that the 
suppression of s-wave fluctuations   is possible to a very high degree
near  half filling,  the commutation relation
$[B,B^\dagger] = L - \hat{N}$ in a sector
$<\hat{N}> \sim \mcl $ , permits both $<B^\dagger B >$ and 
$<B B^\dagger>$ to be simultaneously small.
  This ``squeezing''   results in large
enhancements in the conjugate variables, remarkably enough, without
an appreciable energy cost.  This
effect   has been
termed {\em order by projection}.

While the  arguments  in \cite{shas}
provide a novel direction, the absence of explicit
solutions or of good wave functions has proved to be  a hindrance
in arriving at a thorough
 understanding  of 
the   physics of these models. We provide in this work a wave function which is tractable enough so that  calculations
of all relevant expectation values can be done analytically, which by itself
is rare in many body systems.  Further the wave function catches the essence
of the enhancements mentioned above, and at half filling is argued to
be exact. While we know of no materials to which these models apply,
we make bold to present these models and wave functions,
 since  they address and provide what we believe to be  amongst
the first positive
results\cite{kohn} for 
the important theoretical
question of the possibility of superconductivity without explicit attractive interactions- in fact arising out of pure repulsion.

{\em \bf The model and the variational wavefunction:}
Let us write the Hamiltonian of the model considered in $d$
dimensions as
\beq
 H=T+U_{s}\,B^{\dag }B \;\; . \label{hamiltonian}
\eeq
Here $ B =\sum b_{k}$
is the s-wave Cooper pair operator, and $%
b_k = c_{-k\downarrow }c_{k\uparrow }$ the pair operators, the kinetic energy $T=\sum \epsilon
_{k}n_{k\sigma }$, with the band dispersion $\epsilon _{k}=-2\sum_{\alpha
=1}^{d}\cos (k_{\alpha })$, although later  we will consider a slightly
more general form of the band dispersion in 2-dimensions with orthorhombic
distortion.  The $U_s$ term discourages on site s-wave correlations, and in 
fact projects out s-wave order. In order to get a feeling for its effect, consider the estimate of the energy in a BCS state $|\Psi_{BCS}>= \prod(u_k+
v_k b^\dagger_k) |vac>$. Provided $\sum u_k v_k =0$, we find $E_{BCS}=
\sum [  2 (\epsilon_k - \mu) v_k^2 + U_s v_k^4]$, subject to $N= 2 \sum v_k^2$.
The role of $U_s$ is clearly to flatten out $v_k$ from its step function 
behaviour in the normal state, and hence a superconducting state with true
Long Ranged Order (LRO)  and an extensive energy shift arises,
at least as a  local variational minimum.
In Ref\cite{shas},  a more general model is
introduced including the above term as well as the Hubbard $U$ term. The general
results of \cite{shas} imply for these models that 
away from half filling, the ground state energy 
density\cite{remark} is  unshifted by the $U_s$ term, while the extended s-wave correlations
get a large enhancement at the expense of suppressing s-wave correlations. By 
continuity in density, we expect superconductivity at half filling. In this work
we ignore the Hubbard $U$ term for tractability, and explore the other aspects
within the model of Eq(\ref{hamiltonian}), in particular the origin of the
enhancements.  

The extended s-wave operator 
$A=-2\sum \epsilon _{k}b_{k}$ satisfies the commutation relations $[B,T]=-A$ and $[B,A^{\dagger }]=2T.$
 The main  variational
wave function proposed here is written in terms of the free fermi wave function
 $|\Phi >= \prod_{|k| < k_f} \;  b^{\dagger}_k | vac>$ as
\beq
|\Psi_{\theta }> \;=\exp (-\frac{\theta }{{\cal L}} \; S)\;|\Phi >, \label{wavefn}
\eeq
where the number of lattice sites is  \cl. The prefactor generates 
several Cooper particle pairs
and Cooper hole pairs in the fermi gas. 
To motivate this wavefunction, note that the 
anti Hermitean operator $S= \sum (\ek - \ekp) \bkdag \bkp $ can be viewed
as the commutator $\frac{1}{2} [T, B^\dag B] $, and hence the wave function
may be viewed as a ``rotation'' about a direction orthogonal to the kinetic
and potential energies.
Such  a strategy is familiar e.g. from diagonalizing a
 sum over two Pauli matrices, as 
well as quadratic forms in bosons. A similar rotation also 
gives the exact answer in a 
further simplified model \cite{brijshas}.
 
{\em \bf Calculation of expectation values:}
We now turn to the calculation of expectation values $<Q>_\theta \equiv <\Psi_\theta| Q | \Psi_\theta> $ of various operators $Q$ in the above state.
We begin by evaluating the derivative of the kinetic energy $<T>_\theta$,
\beq
\delth <T>_\theta = \frac{1}{\mcl} [ < A^\dagger A>_\th - 2 <C^\dagger B + B^\dagger C>_\th \; ]  \equiv 4 \alpha(\th) - 2 \nu(\th) , \label{kinetic}
\eeq  
where $C= \sum \epsilon_k^2 b_k$, thus defining the
functions $\alpha, \nu$. We immediately note a feature of considerable importance, namely that unless one of the correlation functions in the right hand side of Eq(\ref{kinetic}) possesses explicit Long Ranged Order (LRO),
the change in the ( extensive)
kinetic energy is {\em not extensive},
and may be neglected in the thermodynamic limit. We will make repeated use of
this insight later. We  also note that this includes the case where
one of the 
correlators possesses Quasi LRO, i.e 
$\sim {\mcl}^{1+\zeta}$, with $\zeta < 1$, indeed we find  later 
 that $\zeta =\frac{1}{2}$  does occur for the case of greatest interest, namely the half filled limit. We also need 
 the expectation of operators 
$T_n \equiv \sum \epsilon_k^n \; \phi_k$, with $\phi_k = n_{k \uparrow} + n_{-k \downarrow} -1 $, 
so that $T_1 = T$. The 
equations are most compactly written in 
terms of the operators $I_n= \sum \epsilon_k^n b_k$  such that $I_0=B ,\; I_1 = - \frhalf A, \; I_2 = C$,
and their correlators $\Gamma_{n,m}(\th) \equiv \frac{1}{\mcl}
< I^\dagger_n I_m + I_m^\dagger I_n >_\th $ as
\beq
\delth < T_n >_\th = 2 [ \Gamma_{n,1}(\th) - \Gamma_{n,0}(\th)] .\label{derivkingen}
\eeq
We note that in view of the above definition,
LRO corresponds to
$\Gamma \sim O({\mcl})$, 
whereas in the absence of true LRO, ~ $\Gamma \sim o({\mcl})$, and hence here
the correlators $< T_n >_\th$ are insensitive to the rotation, at least to leading order in \cl.

The correlators $\Gamma_{n,m}$ satisfy exact equations 
\beq
\delth \Gamma_{n,m}(\th) = \frac{1}{\mcl^2} <[T_{n+1} I_0^\dagger I_m + I^\dagger_n I_0 T_{m+1} - I^\dagger_n I_1 T_m -
 T_n I^\dagger_1 I_m + h.c.]>_\th,
\label{exacteqns}
\eeq
with initial conditions  obtained by
taking the expectation value in the free fermi ground state: they read $\Gamma_{n,m}(0)= 2 \bbl \epsilon_k^{n+m} f_k \bbr$, with  $f_k$ the usual ( non-interacting) fermi function and
$\bbl g_k \bbr \equiv \frac{1}{\mcl} \sum_k g_k$. 

The equations are handled assuming a fundamental factorization that arises in the thermodynamic limit. This
factorization of hermitean global operators ( i.e. sums
over all sites of local operators)  and may be
expressed as the statement that 
for any two such  $Q_j$ with nonzero averages,  we have
 $< Q_1 Q_2>_\th / [<Q_1>_\th <Q_2>_\th ] \sim o(\mcl)$.
This is also equivalent to
the statement that the connected part of the correlator is subleading in powers of $\mcl$. The case needed by us
corresponds to $Q_1 \sim \mcl \; \Gamma_{n,m}$ and $Q_2 \sim T_n$. If this is assumed then we  immediately see that Equations(\ref{exacteqns}) factor out into linear 
equations
\beq
\delth \Gamma_{n,m}= \mu_{n+1} \Gamma_{0,m} +\mu_{m +1} \Gamma_{0,n} -
\mu_m \Gamma_{1,n} - \mu_n \Gamma_{1,m} \label{faceqns}
\eeq
with coefficients $\mu_n \equiv \; \bbl \epsilon_k^n (2 f_k -1) \bbr $
obtainable numerically,
 together with  the initial conditions $\Gamma_{n,m}(0)= 2 \bbl \epsilon_k^{n+m} f_k \bbr $.

We have checked this assumption  to low orders by evaluating the correlation functions $\Gamma_{n,m}$ for the most relevant cases $(n,m)= (0,0), (1,1)$
out to $O(\th^5)$. The calculation proceeds by expanding the expectation value
as a nested commutator with $S$, which is evaluated by Wick's theorem for
ground state correlations,  leading to $n!$ contractions
that are further multiplied by powers of $\epsilon$.
 This results in sums over $k$ of
algebraic functions of the band energies and fermi functions.
 The series agrees with the one generated
from the above factorization procedure exactly to that order, and provides
non trivial support for it.

We now consider the correlation functions of  interest, and begin 
by denoting $\beta(\th)= \frhalf \Gamma_{0,0} , \; \alpha(\th)= \frhalf \Gamma_{1,1} ,
\; \gamma(\th)= \Gamma_{1,0}  $ and finally $\nu(\th)= \Gamma_{2,0} $.   
The equations of  immediate concern are then
\begin{eqnarray}
\delth \alpha & = & - 2 \mu_1 \alpha + \mu_2 \gamma  \nonumber \\ 
\delth \beta & = & 2  \mu_1 \beta + (1- \rho) \gamma  \nonumber \\ 
\delth \gamma & = & 2(1-\rho) \alpha + 2 \mu_2 \beta.  \label{eqns}  
\end{eqnarray}
We also note for later use that
$\delth \nu= \mu_1 \nu + 2 \mu_3 \beta + (1- \rho) \Gamma_{1,2} - \mu_2 \gamma$.  We denote the particle density as $\rho$ such that $\rho \leq 1$.  
The initial conditions are $\beta(0) =\frac{1}{2} \rho  , \; \alpha(0) = \nu(0)=
d + \frac{1}{2} \mu_2, \; \gamma(0) = \mu_1$. We  
now examine  the behaviour of these coupled equations. Let us note
that $\mu_0 = (\rho -1)$, further
 $\mu_1 $ is negative and  non vanishing near half filling, whereas $\mu_2$
also negative,
is very small $\sim O((1-\rho)^3) $ near half filling. In fact for all even $n$
we must have $\mu_n \rightarrow 0$ as we approach half filling, due to particle hole symmetry. If we neglect the cross couplings, then $\beta$ falls off as $\th$ increases from $0$ while 
$\alpha$ increases. 

A neat way to solve these equations is suggested by inspection: we observe
 that in effect the   Eqns(\ref{eqns}) are generated by a linear  relation
$\delth I_n = \mu_{n+1} I_0 - \mu_n I_1$ and the hermitean conjugates of these.
The  $\th$ dependence of all the operators are related thus to
the fundamental ones $I_0, I_1$, and thus we  find linear combinations that 
have simple evolution: $Q_\pm = - \mu_0 I_1 + ( \mu_1 \pm \lambda_0) I_0$ where $\lambda_0 = \sqrt{ \mu_1^2 - \mu_0 \mu_2}$. These obey uncoupled equations $\delth Q_\pm = \pm \lambda_0 Q_\pm$. 

We  readily invert and find with
$\xi_\pm=  ( 1 \pm \frac{\mu_1}{\lambda_0})$, 
$I_0= \frac{1}{ 2 \lambda_0} \{ Q_+ - Q_- \} $ and $ I_1= 
\frac{1}{ 2 (1 - \rho)}  \{ \xi_- Q_+ +
\xi_+ Q_- \} $.  The  four basic correlators $F_{\sigma_1,\sigma_2}(\th)\ =\; <Q^\dagger_{\sigma_1} Q_{\sigma_2} >_\th$
 can thus be found simply:$ F_{\sigma_1,\sigma_2}(\th)= F_{\sigma_1,\sigma_2}(0)
\exp( \lambda_0 \th ( \sigma_1 + \sigma_2) ) $.
The initial conditions are given as
\beq
  F_{\sigma_1,\sigma_2}(0) = (1-\rho)^2 \alpha(0) + (1- \rho) \{ \mu_1 + \frac{\lambda_0}{2} ( \sigma_1 + \sigma_2) \} \gamma(0) + 
(\mu_1 + \sigma_1 \lambda_0) (\mu_1 + \sigma_2 \lambda_0) \beta(0).
\eeq 
We thus find the expectation values:
\begin{eqnarray}
\alpha(\th)& = & \frac{1}{ 4 ( 1 - \rho)^2}[\xi_-^2 F_{1, 1}(0) \exp( 2 \lambda_0 \th) +  \xi_+^2 F_{-1, -1}(0) \exp( - 2 \lambda_0 \th)+  
2 \xi_+ \xi_-   F_{1, -1}(0)] \nonumber \\ 
\beta(\th)& = & \frac{1}{4 \lambda_0^2} [ F_{1, 1}(0) \exp( 2 \lambda_0 \th) + F_{-1, -1}(0)  \exp( - 2 \lambda_0 \th) -  
2    F_{1, -1}(0) ] \nonumber \\ 
\gamma(\th) & = &  \frac{1}{ 2\lambda_0 ( 1 - \rho)}[\xi_- F_{1, 1}(0) \exp( 2 \lambda_0 \th) - 
 \xi_+ F_{-1, -1}(0)\exp( - 2 \lambda_0 \th)+
2 (\frac{\mu_1}{\lambda_0} )F_{1, -1}(0) ] \label{solution}
\end{eqnarray}
These together with Eq(\ref{kinetic}) provide the solution to the variational
problem
since the variational energy
density\cite{remark}  is
$\mu_1 + U_s \beta(\th)$, as well as that of computing the correlations.

We show in Fig(1) the behaviour of the three functions $\alpha, \beta, \gamma$ for the case of
2-d at a typical value of the density $\rho =.75$. Note that  $\beta$
goes through a very shallow minimum with a value $\beta^*(\rho) \equiv \beta(\th^*) \sim 0$,  and $\gamma$ also seems to
go through zero at  nearly the same value of the minimizing $\theta = \th^*$. From the Equations(\ref{eqns}), $\beta^*$ would be actually zero if $\gamma$ vanishes at exactly $\th^*$.
To investigate this further, 
we plot in Fig(2) the minimum value  $\beta^*(\rho)$ as a function of
density in 2-d, and the inset shows a similar plot for 1-d.  We see that for general densities,
 although  $\beta^*  $  is impressively small, it is non zero. From the general arguments
of \cite{shas} we know that the 
exact ground state energy density is unshifted from the non interacting value, corresponding to $\beta_{exact}=0$,  and so our wave function gives a very
 good approximation to the true 
energy for most densities, provided we restrict to $U_s \;\sim |\mu_1|$. In Fig(3) we show   $\alpha^*(\rho)/\alpha(0)$, this is 
enhanced greatly, by  an order of magnitude as far  as $15\%$
away from half filling.
The lower bound 
$\alpha_{lb}= \frac{1}{2} ( \frac{ \mu_1^2}{(1-\rho)} + \mu_2 )$
obtained in  Ref\cite{shas},  also
has a similar behaviour close to half filling.

{\em \bf Solution at Half filling:}
 The discussion of the solution above indicates that for half filling, i.e.
$\rho= 1$, the wave function discussed here could be exact, since $\beta^*$ goes to zero here. Our analysis of the orders of magnitude of the correlators
was based so far on the implicit assumption that $\th \sim O(1)$. From
the solution it can be shown that for $\rho < 1$,
 the minimizing $\th^*\sim -\log(1- \rho)$, and hence at $\rho=1$,
we must work with greater precision to see if
the assumptions made earlier are consistent. Firstly we  assume that Eqns(\ref{eqns}) are still valid
at half filling, requiring that $\lim_{\mcl \rightarrow \infty} <T_n>_\th/ \mcl$ is unchanged from its non-interacting value $\mu_n$. We verify
at the end that this is still true and hence obtain a self consistent argument. With the assumption
we  set  $\mu_{2n} \rightarrow 0$ and hence the equations decouple and
 are integrated immediately giving $\alpha(\th) = \alpha(0) \exp( - 2 \mu_1 \th)$ and $\beta(\th)= \beta(0) 
\exp( 2 \mu_1 \th)$. We  also integrate the equation for $\nu$ and find $\nu(\th)= g_1 \exp( \mu_1 \th)+
g_2 \exp( 2 \mu_1 \th)$. The constants $g_1, g_2$ involve the initial conditions, but are not important
since we see that for large (positive) $\th$ we can safely neglect $\nu(\th)$ ( recall that 
$\mu_1 < 0$). We substitute for $\alpha$ into Eqn(\ref{kinetic}) and integrate over $\th$, to find the energy per site\cite{remark} 
\beq
e(\th) = e_{non} + \frac{U_s}{2} \exp( 2 \mu_1 \th)  + \frac{2 \alpha(0)}{ |\mu_1| \mcl} \exp( - 2 \mu_1 \th) +
\mbox{ (negligible terms)} \label{enerhalf}
\eeq
where $e_{non}= \mu_1$ is the 
ground state energy density  of the free fermi gas, and
 the negligible terms  are of the type $\frac{1}{\mcl} \{ 1, \exp( \mu_1 \th), \exp( 2 \mu_1 \th) \} $
etc. We see that minimizing Eq(\ref{enerhalf}) w.r.t. $\th$ pushes up the energy as
\beq
e(\th^*)= e_{non} + 2 \frac{\sqrt{U_s \alpha(0) }}{\sqrt{|\mu_1| \mcl}} +
 \mbox{ (negligible terms)}        \label{ernerhalfinal}
\eeq
where $\th^* =1/( 4  |\mu_1|)\;\log( U_s |\mu_1| \mcl /4 \alpha(0)) $. The energy density continues to be  independent of $U_s$.
Thus the growth of $\th$  stops on the scale
of $\log(1/\mcl)$, and we find $\alpha^* = \frac{1}{2}  \sqrt{\alpha(0) U_s \mcl |\mu_1|}$. Since
$\alpha^*$ does not have true LRO, our original assumption is validated, and hence we have a consistent
solution. The solution at half filling thus has Quasi LRO since $<A^\dagger A> \sim
\mcl^{3/2}$, and is superconducting in the sense of having power law order,
as e.g. in a 2-d superconductor at finite temperatures.

{\em \bf The d-wave superconductor:}
We now ask the question whether this class of wave functions works for d-wave superconducting correlations
as well in 2-d. This case is clearly of topical interest in the context of the 2-d High $T_c$ cuprates.
 There is one possibility inherent in the discussion of Ref(\cite{shas}), wherein we take
$\hat{B}= \sum \eta_k b_k$, with $\eta_k= \sign(w_k)$, and $w_k=\cos(k_x ) - \cos( k_y)$. This choice
of the relative phases of the $k$ state Cooper pairs preserves the important commutation relation
$[\hat{B}, \hat{B}^\dagger]= \mcl - \hat{N}$ (where $\hat{N}$ is
the number operator), since $\eta_k^2=1$, and all the arguments of
Ref(\cite{shas}) can be taken over, exactly as for the case of  extended s-wave pairing discussed above.
The conjugate variable that is enhanced now is $\hat{A}= - 2 \sum \epsilon_k \eta_k b_k$, and the
two particle wave function has the signs that  are usually associated with d-wave pairing.
The ``gap'' within a weak coupling approach in this case may be taken to be $\Delta = \Delta_0 \epsilon_k
\eta_k$, and thus has jumps in $k$ space where one expects nodes. Of course this may not be a fatal
difficulty since there is no clear connection between the ground state obtained within the present 
framework, and the low lying excitation spectrum within a BCS framework.

Within the variational framework, we can admit d-wave pairing provided the lattice has a nonzero 
orthorhombic distortion, i.e. the band dispersion is asymmetric in $x$ and $y$ directions. We can 
model this by setting $\epsilon_k = -2 (1- \epsilon_0) \cos(k_x) - 2 ( 1 + \epsilon_0) \cos(k_y)$ with
$\epsilon_0$ a measure of the orthorhombicity. We  choose a variational wave function exactly as
in Eq(\ref{wavefn}), with $S_{DW}= \sum (w_{k_1} - w_{k_2} )\bkdag \bkp$ and define the d-wave pairing
operator $D= \sum w_k b_k$.  With this we can essentially borrow all the results of the previous
calculation provided we replace $ \mu_n \rightarrow \nu_n$, where $\nu_n = \frac{1}{\mcl} < \hat{T}_n >_0$
and $\hat{T}_n= \sum w_k^n \phi_k$. The fermi groundstate has the symmetry of the $\epsilon_k$
and hence we see 
the need for the distortion, for in its  absence,
 $\nu_n =0$ for all $n \geq 1$.
We must also replace the initial conditions appropriately, thus $\Delta(\th) \; = \frac{1}{\mcl} < D^\dagger
D>_\th$, and $\Delta(0) = \bbl w_k^2 f_k \bbr$.  We need to examine the reality of  $\hat{\lambda}_0= \sqrt{ \nu_1^2 + ( 1 - \rho) \nu_2}$.
 In case
$\hat{\lambda}_0$ is imaginary, the nature of solution changes drastically, and we have 
a limited enhancement, since the functions now oscillate as trigonometric functions, as $\th$ varies.
If $\hat{\lambda}_0$ is real,
 we can repeat everything said before and qualitatively
find the same answers as for
the extended s-wave case. The numerics require a knowledge of the elements $\nu_n$ and we will not 
discuss them further here.
 At half filling the state is similarly obtained, and the energy is 
different from that of the s-wave case only in terms of $O(1/\sqrt{\mcl})$, it is 
higher in general since $\nu_n $ are usually smaller than the $\mu_n$ for $n \geq 1$.
It must be remarked that in this case the correlation function $\alpha$ is also enhanced 
similarly to $\Delta$,  since these correlations get coupled. The correct symmetry of the order parameter
is then of a mixed s and d type, rather than pure s or d\cite{leggett}, and indeed one should, in principle,
optimize the form of the $S$ operator  by taking a 
suitable mixture of s and d functions  to get the best possible energy.

{\em \bf  Summary and Discussion:}
In summary, we have presented  variational wave functions that seem
interesting on several counts.
In many regards, we are tempted to say that the wave functions 
are at least as interesting as the models if not more.
 Firstly, the expectation values are computable
analytically and throw light on an interesting and novel relationship, namely
the enhancement in certain correlations at the expense of on site s-wave correlations. At half filling we have a state that is degenerate with
the free fermi gas in the energy-density, and yet has critical superconducting
 correlations. Is this
 state the exact ground state of the model in Eq(\ref{hamiltonian})?
 The coincidence of the 
{\em energy density} with the
obvious lower bound $e_{non}$, is not a clinching  argument  by itself, particularly
in view of the main message  of our work. The nature of the correlations,
on the other hand, 
being  of the form expected from the uncertainty principle arguments, provide
a significant positive indicator. 
We feel that certain results found here are likely
to be true, such as $\alpha^* \propto \mcl^{1/2} $, and the form of energy corrections
as in  Eq(\ref{ernerhalfinal}),
without however, being able to prove them in a 
rigorous fashion.  Of the values of the coefficients of the leading behaviour, it  is less easy to be
certain.

 The critical behaviour
of the Cooper correlation function
 $<A^\dagger A>$  is  of particular interest. It  behaves as $\mcl^{\frac{3}{2}}$ at $\rho=1$, and  is $\propto \frac{\mcl}{1 - \rho}$
otherwise, suggesting that the  local 
density $\alpha(r)$ of the $A$ operator  has correlations $<\alpha^\dagger(r)
\alpha(0)> \sim \frac{1}{|r|^{d/2}} h(r/\xi(\rho))$, with a density dependent
correlation length $\xi(\rho) \sim \frac{1}{(1-\rho)^{2/d}}$, that diverges
at half filling. Such a crossover is
similar
to that in 
the behaviour of the spin correlations in the Hubbard model in 1-d near
half filling\cite{imada}.
Note that the behaviour of this function for the free fermi gas
is $\sim \frac{1}{r^{2d}}$, and hence the correlations in this 
case  are considerably
shorter  ranged than ours. 

We conclude by mentioning
another nontrivial  implication of the  ideas discussed above, but
going beyond the context of the models considered here.  We have seen that the fermi gas on a lattice
has the possibility of novel large enhancements in the ground
state  correlation functions of Cooper pairs, produced by repulsive terms
of certain kinds. 
These are not reflected in either 
the  total energy (and hence the compressibility) or the momentum distribution 
function. The latter
can be readily seen to be  as sharp as in the ideal fermi case.
Similarly one can check that the spin correlations are unaffected.
The usual fermi liquid enhancements  in spin and charge response that herald incipient
instabilities  are  totally missing in these  systems.
As a result, any bosonic excitations available, including phonons,
would find the electronic system much more prone to superconductivity
of the usual sort than expected on grounds
of the missing fermi liquid enhancements, and presumably be accompanied
by enhanced $T_c$'s.  
Estimates of such effects must await
information on the   spectral weight in the dynamical Cooper susceptibility.

\begin{flushleft}
{\bf Acknowledgements}
\end{flushleft}
I am grateful to several colleagues at Bangalore for useful remarks. This                                                    
work is partially supported by a grant from the Department of
Science and Technology (DSTPHY/AKS/380) and the JNCASR.


\newpage


\begin{figure}
  \begin{center}
  \leavevmode
  \epsfxsize=9.5cm
  \epsfbox{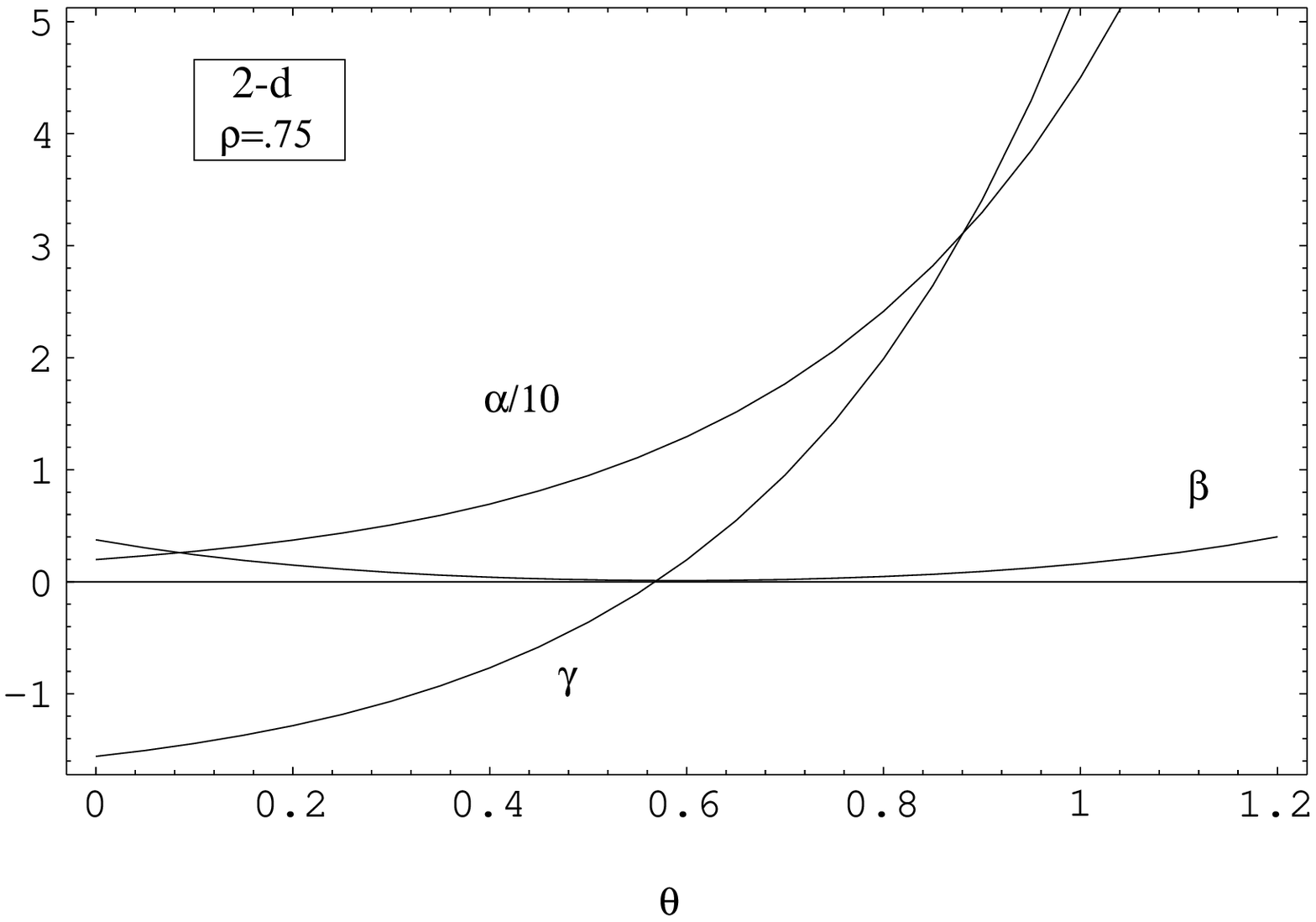}
  \end{center}
  \caption{ The three correlations functions at $\rho=0.75$ in 2-d vs. $\th$.  \label{fig1}}
\end{figure}

\begin{figure}
  \begin{center}
  \leavevmode
  \epsfxsize=9.5cm
  \epsfbox{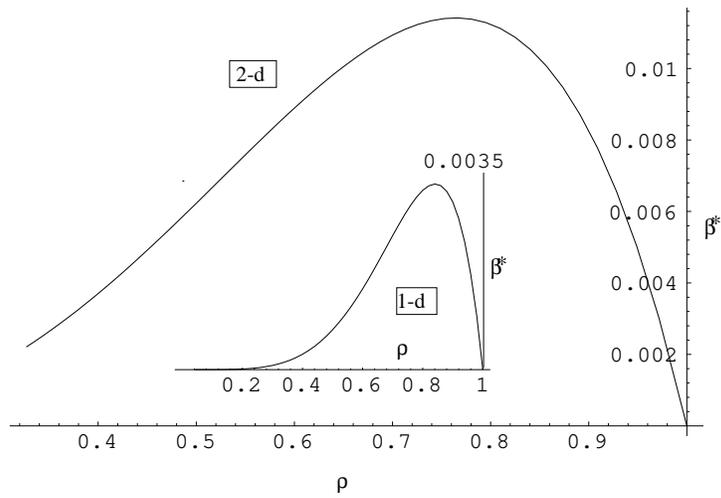}
  \end{center}
  \caption{  The minimum value of $\beta^*$ vs density. \label{fig2}}
\end{figure}

\begin{figure}
  \begin{center}
  \leavevmode
  \epsfxsize=9.5cm
  \epsfbox{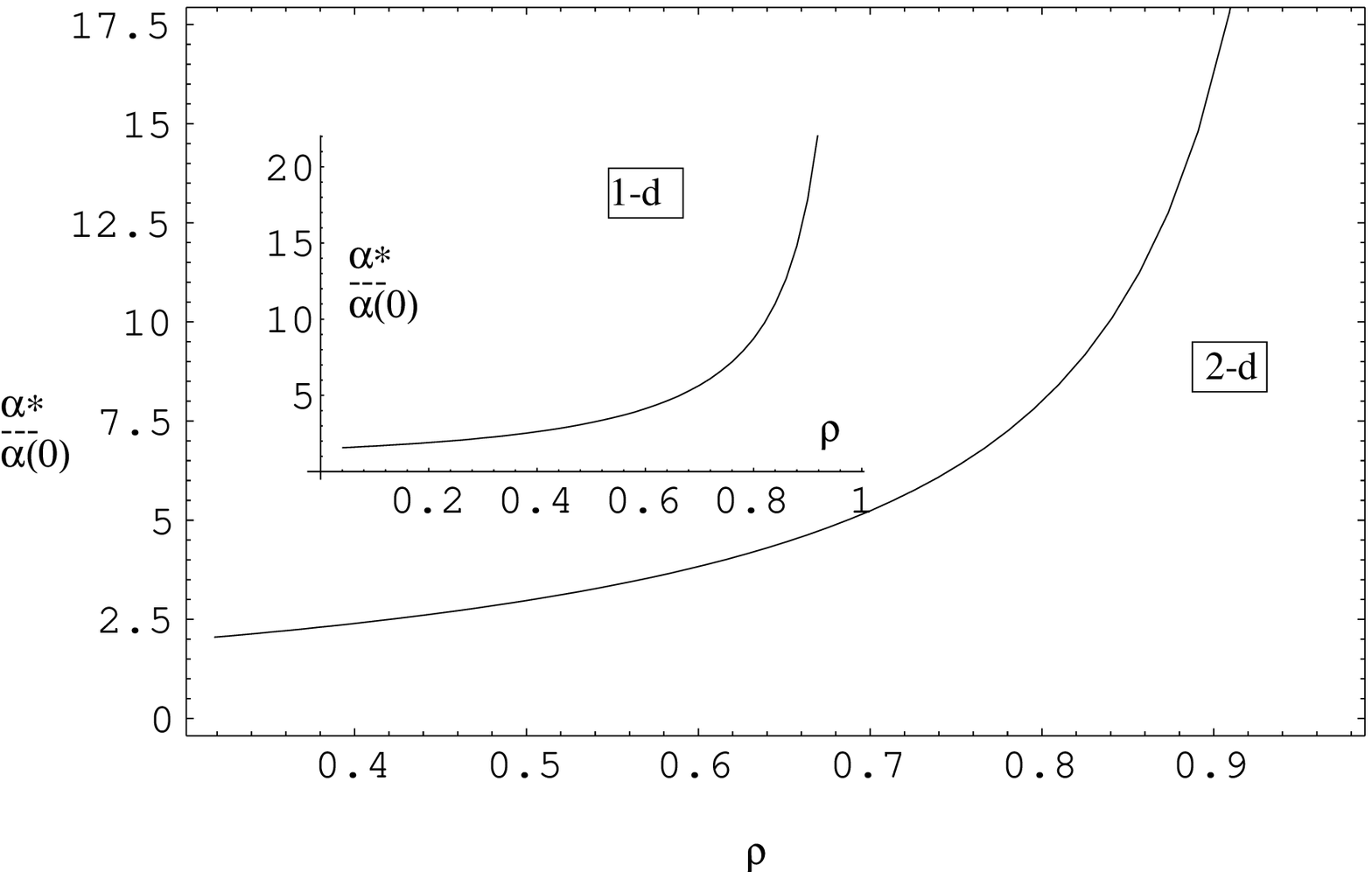}
  \end{center}
  \caption{ The enhancement factor of the
 extended s-wave correlator $\alpha^*/\alpha(0)$  vs density
in 2-d, and in 1-d (inset).
  \label{fig3}}
\end{figure}


\begin{thebibliography}{99}
\bibitem{remark}  The term ``energy density'' is used 
 to denote $\lim_{\mcl -> \infty}
\frac{E_0}{\mcl}$. We use the  term ``energy per site'', on the other hand, to
denote $  \frac{E_0}{\mcl}$ preparatory to taking the thermodynamic limit,
and display only $\th$ dependent
corrections such as $\frac{1}{\mcl^\alpha} g(\th)$. We omit displaying other
finite size corrections that are undoubtedly present but uninteresting, being
$\th$ independent. We occasionally also omit displaying $\th$ dependent but
smaller corrections than the ones displayed.
\end{thebibliography}
\end{document}